\documentclass[10pt,aps,prd,nofootinbib,reprint]{revtex4-1}

\usepackage[utf8]{inputenc}
\usepackage{amsmath}
\usepackage{graphicx}
\usepackage{multirow}

\usepackage[colorlinks,pdfusetitle]{hyperref}
\hypersetup{colorlinks=true,allcolors=[rgb]{1,0.56,0}}

\newcommand{\orcid}[1]{\href{https://orcid.org/#1}{#1}}

\newcommand{\eps}{\epsilon}
\begin{document}

\title{New reactor data improves robustness of neutrino mass ordering determination}

\author{Peter B.~Denton}
\email{pdenton@bnl.gov}
\thanks{\orcid{0000-0002-5209-872X}}
\affiliation{High Energy Theory Group, Physics Department, Brookhaven National Laboratory, Upton, NY 11973, USA}

\author{Julia Gehrlein}
\email{jgehrlein@bnl.gov}
\thanks{\orcid{0000-0002-1235-0505}}
\affiliation{High Energy Theory Group, Physics Department, Brookhaven National Laboratory, Upton, NY 11973, USA}

\begin{abstract}
In neutrino oscillation physics numerous exact degeneracies exist under the name LMA-Dark.
These degeneracies make it impossible to determine the sign of $\Delta m^2_{31}$ known as the atmospheric mass ordering with oscillation experiments alone in the presence of new neutrino interactions.
The combination of different measurements including multiple oscillation channels and neutrino scattering experiments lifts some aspects of these degeneracies.
In fact, previous measurements of coherent elastic neutrino nucleus scattering (CEvNS) by COHERENT already ruled out the LMA-Dark solution for new physics with mediators heavier than $M_{Z'}\sim50$ MeV while cosmological considerations disfavor these scenarios for mediators lighter than $M_{Z'}\sim3$ MeV.
Here we leverage new data from the Dresden-II experiment which provides the strongest bounds on CEvNS with reactor neutrinos to date.
We show that this data completely removes the degeneracies in the $\nu_e$ sector for mediators down to the MeV scale at which point constraints from the early universe take over.
While the LMA-Dark degeneracy is lifted in the $\nu_e$ sector, it can still be restored in the $\nu_\mu$ and $\nu_\tau$ sector or with very specific couplings to up and down quarks, and we speculate on a path forward.
\end{abstract}

\date{July 22, 2022}

\maketitle

\section{Introduction}
The question about the sign of $\Delta m^2_{31}$, the atmospheric neutrino mass ordering -- that is, whether the mass eigenstate that has almost no $\nu_e$ is the lightest or the heaviest of the three -- is one of the few remaining unknowns in the Standard Model (SM) of particle physics.
A positive value of $\Delta m^2_{31}$ results in a larger $\nu_\mu\to\nu_e$ appearance probability and a smaller $\bar\nu_\mu\to\bar\nu_e$ appearance probability at the first oscillation maximum in matter.
In addition, this open question has additional important phenomenological implications in many areas beyond oscillation physics, such as neutrinoless double-beta decay searches, beta-decay end point mass measurements, cosmological measurements of neutrino masses, and the cosmic neutrino background. 
While analyses of oscillation data have suggested that the mass ordering is normal at the $\sim3\sigma$ level \cite{deSalas:2018bym,deSalas:2020pgw,Esteban:2020cvm}, this story has been somewhat complicated by the latest NOvA and T2K data \cite{NOvA:2021nfi,T2K:2021xwb,Kelly:2020fkv,Denton:2020uda,Chatterjee:2020kkm} which have weakened the evidence for the normal ordering.

In order to truly understand the neutrino oscillation picture, one must consider many new physics scenarios that modify oscillations \cite{Arguelles:2019xgp,Arguelles:2022xxa}.
The possible presence of new physics in the neutrino sector, such as neutrino non-standard interactions (NSIs) \cite{Wolfenstein:1977ue,Dev:2019anc} makes it impossible to determine the mass ordering with oscillation experiments alone.
NSIs are a general effective field theory framework of new interactions between neutrinos and matter fermions such as up quarks, down quarks, and electrons.
These interactions act as an additional matter effect in a new basis leading to distinct flavor effects.
Numerous UV complete models with sizable NSI exist in the literature \cite{Forero:2016ghr,Denton:2018dqq,Dey:2018yht,Babu:2017olk,Farzan:2016wym,Farzan:2015hkd,Farzan:2015doa,Babu:2019mfe}.

The difficulty in determining the mass ordering is due to several exact degeneracies in the neutrino oscillation Hamiltonian which are often grouped under the name LMA-Dark\footnote{The name LMA-Dark comes from the Large Mixing Angle (LMA) solution to the solar neutrino problem which is now known to be the correct solution with $\theta_{12}\sim33^\circ$.
The term dark refers to the ``dark'' side of the solution; either that $\theta_{12}=57^\circ$ or that $\Delta m^2_{21}<0$, depending on one's choice of definition; see \cite{Denton:2020exu,Denton:2021vtf}.} \cite{deGouvea:2000pqg,Bakhti:2014pva,Coloma:2016gei,Denton:2021vtf}.
Simply put, neutrino oscillations are unchanged when adding new physics that changes the sign of the matter effect, and simultaneously swapping the atmospheric mass ordering.
That is, 
\begin{align*}
P_{\alpha\beta}({\rm NO},L,E,\rho,\eps=0)&=P_{\alpha\beta}({\rm IO},L,E,\rho,\eps=-2)\,,\\
P_{\alpha\beta}({\rm IO},L,E,\rho,\eps=0)&=P_{\alpha\beta}({\rm NO},L,E,\rho,\eps=-2)\,,
\end{align*}
where $P_{\alpha\beta}$ is any oscillation probability from flavor $\alpha$ to flavor $\beta$ for either neutrinos or anti-neutrinos, NO (IO) refers to the normal (inverted) atmospheric mass ordering, $L$ is the baseline, $E$ is the neutrino energy, $\rho$ is the density of matter the neutrino is propagating through, and $\eps$ is a particular new physics parameter to be discussed in section \ref{sec:lmad} below.
No other new physics scenario leads to this \emph{exact} degeneracy among all possible oscillation experiments.

Hence the presence of new physics leads to an ambiguity in the determination of the neutrino mass ordering.
As oscillations are unable to identify the new physics associated to LMA-Dark, one must look elsewhere, notably to scattering experiments, for which the NSI aspect of the LMA-Dark solution can be ruled out, but only under certain assumptions \cite{Coloma:2017egw,Coloma:2017ncl,Denton:2018xmq}. 
Since oscillation experiments are only sensitive to forward elastic scattering effects modifying the potential, they can only constrain $\eps\propto g^2/M_{Z'}^2$ and are not sensitive to the mediator mass\footnote{This statement does not apply to very light mediators ($M_{Z'}\sim\mathcal{O}(10^{-11})$ eV) whose wavelength is of the order of the depth the neutrinos penetrate the Earth \cite{Grifols:2003gy,Joshipura:2003jh,Davoudiasl:2011sz,Wise:2018rnb,Smirnov:2019cae,Babu:2019iml,Coloma:2020gfv}. In this case the neutrinos feel a depth dependent potential due to the new mediator.}.
Thus, from oscillation data alone, the degeneracy persists for all mediator masses.
Neutrino scattering experiments do not have the same degeneracy that oscillations do, but scattering experiments are only sensitive to NSI with sufficiently heavy mediator masses, $M_{Z'}^2\gtrsim q^2$ where $q^2$ is the square of the momentum transfer of the detection process.
Thus we look to the lowest threshold neutrino scattering experiments to provide constraints on LMA-Dark.

Coherent elastic neutrino nucleus scattering (CEvNS) is an essentially thresholdless channel which has been detected in low threshold experiments.
The COHERENT experiment was the first to detect CEvNS \cite{COHERENT:2017ipa}. It makes use of neutrinos from pion decay at rest and probes LMA-Dark with mediator masses down to $M_{Z'}\sim50$ MeV \cite{Coloma:2017egw,Denton:2018xmq}.

In addition to neutrino scattering experiments, information about the early universe provides a probe of new light particles.
Data from the early universe conservatively disfavors mediators lighter than $\sim3$ MeV \cite{Kamada:2015era,Huang:2017egl,Blinov:2019gcj,Sabti:2021reh}.
This leads to a currently allowed range of mediator masses of $3~\text{MeV}\lesssim M_{Z'}\lesssim 50~\text{MeV}$ for which large NSI of a comparable size to the weak interaction is still viable.
This region can be probed by measuring CEvNS with a very low threshold detector at a reactor neutrino experiment \cite{Denton:2018xmq}, a measurement a number of experiments are working on \cite{chillax,CONNIE:2019swq,CONUS:2020skt,MINER:2016igy,nugen,Ni:2021mwa,NUCLEUS:2019igx,Giampa:2021wte,Wong:2015kgl,NEWS-G:2017pxg,Ricochet:2021rjo,Akimov:2017hee,Colaresi:2022obx,Choi:2022nxu}.
In this paper we will mostly remedy the deficiency of unprobed LMA-Dark parameter space by analyzing the latest data from the Dresden-II experiment \cite{Colaresi:2022obx}, the first experiment that has evidence for CEvNS with reactor neutrinos, in the context of this degeneracy.

We begin this paper by discussing in detail the degeneracies related to neutrino oscillations and how they can be broken in section \ref{sec:lmad}.
Next, we discuss the recent Dresden-II data in section \ref{sec:dresden} and show our numerical results in section \ref{sec:results}.
Finally, we discuss likely future improvements in section \ref{sec:future} and conclude the paper in section \ref{sec:conclusions}.

\section{LMA-Dark Degeneracies}
\label{sec:lmad}
There are a number of degeneracies related to what is known as LMA-Dark, which leads to some confusion in the literature.
These degeneracies are exact in some contexts, easily broken in some contexts, and only softly lifted in others.
The different degeneracies depend on whether oscillations or a scattering process is considered, whether the new physics depends on the mediator mass or not, whether it depends on the specific quark couplings or not, and whether it is in the $\nu_e$ sector or the $\nu_\mu$ and $\nu_\tau$ sectors.
We will first review them and present the entire picture here for clarity.
To begin, we write down the Hamiltonian governing neutrino oscillations in the presence of NSI,
\begin{multline}
H=\frac1{2E}\left[U
\begin{pmatrix}
0&0&0\\
0&\Delta m^2_{21}&0\\
0&0&\Delta m^2_{31}
\end{pmatrix}
U^\dagger+\right.\\
\left.a
\begin{pmatrix}
1+\eps_{ee}&\eps_{e\mu}&\eps_{e\tau}\\
\eps_{e\mu}^*&\eps_{\mu\mu}&\eps_{\mu\tau}\\
\eps_{e\tau}^*&\eps_{\mu\tau}^*&\eps_{\tau\tau}
\end{pmatrix}
\right]
\,,
\end{multline}
where $E$ is the neutrino energy, $U$ is the lepton mixing matrix \cite{Pontecorvo:1957cp,Maki:1962mu}, $a\equiv2\sqrt2G_FN_eE$ is the matter potential, $G_F$ is Fermi's constant, $N_e$ is the electron number density, and the $\eps_{\alpha\beta}$ terms quantify the magnitude of the NSI relative to $G_F$, and the 1 next to $\eps_{ee}$ is due to the SM matter effect.
The Hamiltonian level NSI parameters for oscillations are related to the Lagrangian level NSI parameters via
\begin{equation}
\eps_{\alpha\beta}=\sum_{f\in\{e,u,d\}}\frac{N_f}{N_e}\eps_{\alpha\beta}^{f,V}\,,
\end{equation}
with the fermion matter density $N_f$.
The typical effective field theory description of NSI is
\begin{equation}
\mathcal L_{\rm NSI}=-2\sqrt2G_F\sum_{f,\alpha,\beta}\eps^{f,V}_{\alpha\beta}(\bar\nu_\alpha\gamma^\mu P_L\nu_\beta)(\bar f\gamma_\mu f)\,,
\end{equation}
where we have only included the vector component.
Note that we will differentiate between Lagrangian and Hamiltonian level NSI as described above by the presence of a superscript or lack thereof, respectively.
For oscillations, it is useful to define $Y_n\equiv N_n/N_e$ which is 1.05 in the Earth and 0.30 in the Sun for the peak of the $^8$B flux from the BS05(OP) standard solar model \cite{Bahcall:2004pz}.
In the following, we will focus on the diagonal NSI parameters only; we will discuss the off-diagonal parameters at the end of this section.

\medskip

\begin{figure}
\centering
\includegraphics[width=\columnwidth]{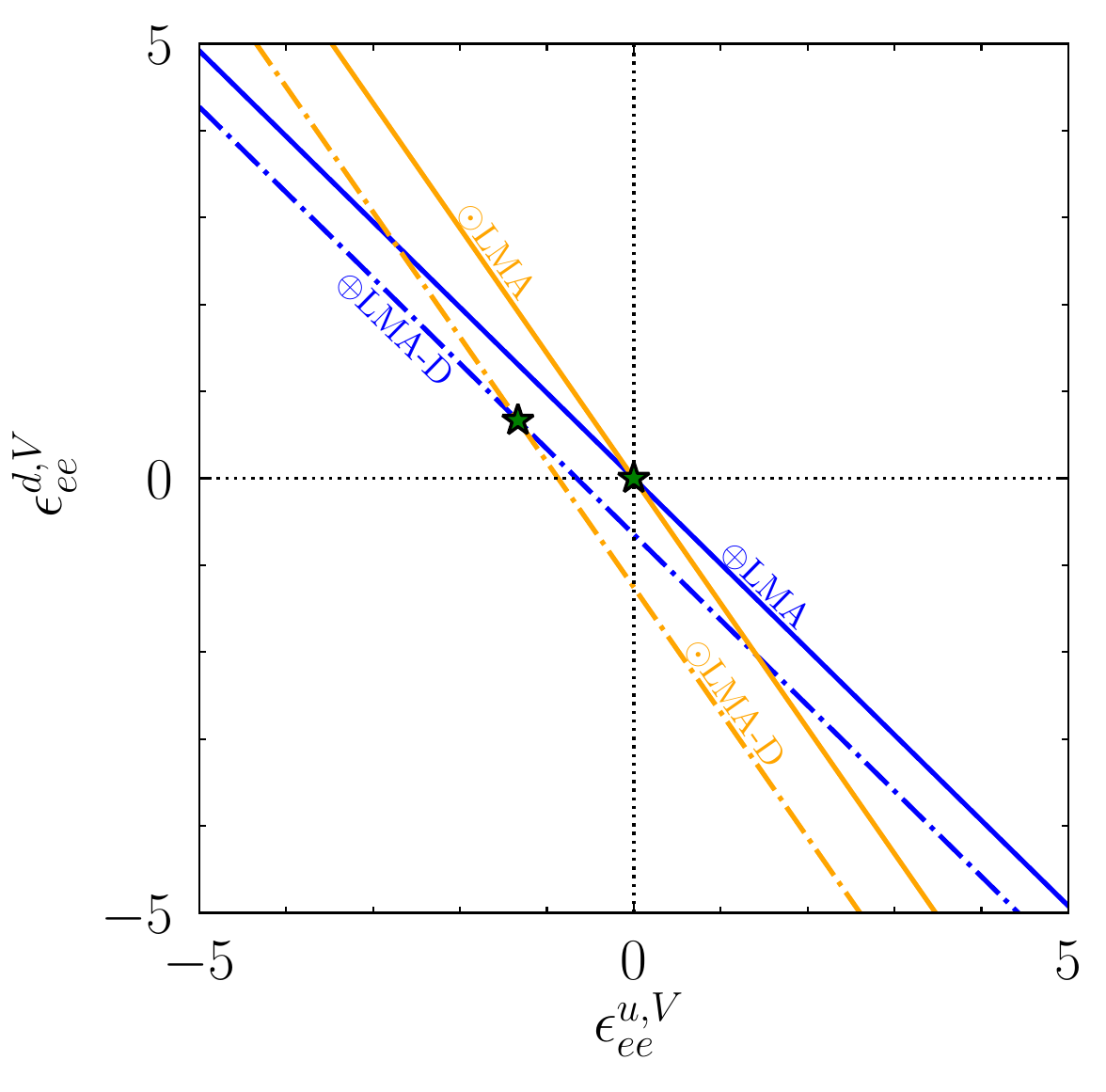}
\caption{The parameter space of the various oscillation degeneracies in terms of couplings to electron neutrinos and up and down quarks using the matter densities of the Earth and the Sun.
The two stars highlight the cases of no new physics and the point that is fully degenerate among oscillation experiments in any material.}
\label{fig:degeneracies}
\end{figure}

Now we walk through the various degeneracies, how they are broken, and how they are restored again, assuming that experiments measure the SM.
\begin{enumerate}
\item 
There is an exact degeneracy without new physics referred to as the dark side\footnote{This is what spawned the name LMA-Dark.} relevant for \textbf{vacuum oscillations} \cite{deGouvea:2000pqg}.
That is, all oscillations in vacuum cannot distinguish between a given scenario, and another with\footnote{Here we take the definition of the mass eigenstates based on $|U_{e1}|>|U_{e2}|>|U_{e3}|$.
This leads to $\theta_{12}<45^\circ$ by definition with the sign of $\Delta m^2_{21}$ to be determined experimentally.
Another common definition of the mass eigenstates is $m_1<m_2$, $|U_{e1}|>|U_{e3}|$, and $|U_{e2}|>|U_{e3}|$, in which case $\Delta m^2_{21}>0$ by definition and the octant of $\theta_{12}$ is to be determined experimentally.
Different definitions of the mass eigenstates lead to different definitions of the LMA-Dark degeneracy.
For an overview of these definitions, see \cite{Denton:2020exu,Denton:2021vtf}.}
\begin{equation}
\begin{gathered}
\Delta m^2_{21}\to-\Delta m^2_{21}\,,\quad
\Delta m^2_{31}\to-\Delta m^2_{31}\,,\\
\delta\to-\delta\,,
\end{gathered}
\end{equation}
which is equivalent to sending the Hamiltonian $H_{\rm vac}\to-H_{\rm vac}^*$ which should leave all physical observables unchanged, assuming CPT invariance.
\item
Measurements of neutrino \textbf{oscillations in matter} break this degeneracy.
The Hamiltonian has the matter potential as an additional term which does not change sign along with the vacuum oscillation parameters as described above.
Therefore, the agreement of the results from SNO (in matter) \cite{Ahmad:2002jz} and of the results from KamLAND (in vacuum) \cite{Gando:2013nba} or pp solar oscillations (in vacuum) \cite{BOREXINO:2018ohr,SAGE:1994ctc,GALLEX:1992gcp} breaks this degeneracy.
It is only with the measurement of solar parameters in the dense solar environment at SNO that JUNO \cite{Djurcic:2015vqa} or a combination of JUNO and upcoming atmospheric disappearance data \cite{IceCube-Gen2:2019fet,KM3NeT:2021rkn} can measure the atmospheric mass ordering \cite{Nunokawa:2005nx} without an independent measurement of the atmospheric mass ordering in matter; see the appendix for more discussion on JUNO.
Alternatively, DUNE (in matter) \cite{DUNE:2020ypp} can determine the atmospheric mass ordering by combining the appearance channel with the disappearance channel's measurement of $\Delta m^2_{31}$ which is essentially unaffected by matter, or any other measurement of $\Delta m^2_{31}$.
\item
In the presence of \textbf{NSI}, the \textbf{degeneracy is exactly restored for oscillations} \cite{Bakhti:2014pva,Coloma:2016gei,Miranda:2004nb}.
In the SM the matter effect does not change sign under CPT, but a value of $\eps_{ee}=-2$ is exactly equivalent to the necessary sign change.
The other diagonal NSI parameters must be zero at this point.
\item
For a given set of couplings to up quarks and down quarks, precise measurements of $\eps_{ee}$ in \textbf{different materials}, such as the Sun and the crust of the Earth, can lift the degeneracy.
\item
If the degeneracy at $\eps_{ee}=-2$ is present and the new mediator is coupled to electrons or to a specific combination of up and down quarks: $\eps_{ee}^{u,V}=-4/3$ and $\eps_{ee}^{d,V}=2/3$, then \textbf{no combination of oscillation experiments} can break the degeneracy, see fig.~\ref{fig:degeneracies}.
\item
Neutrino \textbf{scattering} experiments such as NuTeV, CHARM, COHERENT, and Dresden-II (discussed here in this context for the first time) do not suffer from the LMA-Dark oscillation degeneracy and therefore they provide a probe of the NSI part of the LMA-Dark parameter space \cite{Coloma:2017egw,Coloma:2017ncl,Denton:2018xmq}.
\item
Constraints on LMA-Dark from scattering experiments only apply if the momentum transfer in the new interaction
is larger than the minimal energy measurable by the experiment in question \cite{Coloma:2017egw,Liao:2017uzy,Denton:2018xmq}, this puts a lower bound on the mediator mass which can be probed.
Previously, COHERENT provided the lowest threshold constraint on the $\eps_{ee}$ part of LMA-Dark down to mediator masses $\sim50$ MeV, for all relevant combinations of $\eps_{ee}^{d,V}-\eps_{ee}^{u,V}$ \cite{Denton:2018xmq,Chaves:2021pey},  while new data from Dresden-II presented in this paper constrains the $\eps_{ee}$ part of LMA-Dark down to $\sim1$ MeV.
Scenarios with \textbf{lighter mediators} are not constrained by scattering.
\item
\textbf{Early universe measurements constrain mediators} lighter than $\sim3$ MeV for couplings corresponding to LMA-Dark \cite{Kamada:2015era,Huang:2017egl,Sabti:2021reh}.
Provided that scattering experiments can reach this scale, then the degeneracy can be lifted for any mediator mass.
\item
The \textbf{diagonal degeneracy} of the oscillation Hamiltonian allows one to slide the oscillation degeneracy into the $\eps_{\mu\mu}$ and $\eps_{\tau\tau}$ terms \cite{Denton:2018xmq} as one can subtract one diagonal element of the Hamiltonian, making a measurement of $\eps_{ee}$ from Dresden-II or any reactor neutrino experiment, powerless which leads to the COHERENT data being the most sensitive to constrain this case.
That is, $\eps_{ee}=-2$ is equivalent, at oscillation experiments, to $\eps_{ee}=0$ and $\eps_{\mu\mu}=\eps_{\tau\tau}=+2$. This can be parameterized by $x$ via
\begin{equation}
(\eps_{ee},~\eps_{\mu\mu},~\eps_{\tau\tau})=(x-2,~x,~x)\,.
\label{eq:x}
\end{equation}
\item As oscillation experiments are only sensitive to the difference $\eps_{\mu\mu}-\eps_{\tau\tau}$, to rule out LMA-Dark in the muon and tau sector one is required to{\textbf{ constrain $\epsilon_{\mu\mu}=\epsilon_{\tau\tau}=2$ for small mediator masses}} down to $M_{Z'}\gtrsim 3$ MeV in scattering experiments. COHERENT already rules out the LMA-Dark solution for any values of $x$ from Eq.~(\ref{eq:x}) for $M_{Z'}\gtrsim 50$ MeV \cite{Denton:2020hop}.
There is no source of low-energy tau neutrinos at scattering experiments which allows to probe $\eps_{\tau\tau}$.
\item
A \textbf{low-threshold} $\boldsymbol{\pi}$\textbf{-DAR} measurement of CEvNS such as with the Coherent CAPTAIN Mills experiment \cite{CCM:2021leg} or a CEvNS experiment at the European Spallation Source \cite{Baxter:2019mcx, Chaves:2021pey} could probe $\eps_{\mu\mu}$ for low enough mediator masses and therefore cover all of the above cases \cite{Shoemaker:2021hvm}.
\end{enumerate}

The current status of the LMA-Dark degeneracy is that, with this current paper, all of the above steps are satisfied through point number 8 above and thus the degeneracy sits at point number 9: $\eps_{\mu\mu}=\eps_{\tau\tau}=2$ is allowed for some mediator masses in the range $3$ MeV $\lesssim M_{Z'}\lesssim40$ MeV.
In addition, not all of the degeneracy discussed in point number 5 above is lifted; that is certain combinations of $\eps_{ee}^{u,V}$ and $\eps_{ee}^{d,V}$ are still allowed at $M_{Z'}\sim3$ MeV.

There are several caveats to this narrative.
In point number 2 above, we note that it may be possible for DUNE to measure the mass ordering using the matter effect with the neutrino and anti-neutrino appearance channels \emph{alone} depending on the quality of the measurement across the first oscillation maximum.

In point number 5 we mention that the new interaction can be with electrons in addition to up quarks or down quarks.
NSI with electrons required for the LMA-Dark degeneracy is constrained by Borexino \cite{Dutta:2020che,BOREXINO:2018ohr} as well as TEXONO, CHARM-II, and GEMMA data \cite{Lindner:2018kjo,TEXONO:2009knm,CHARM-II:1994dzw,Beda:2010hk} at a level low enough to reach the cosmology constraints.
If the coupling to electrons is only with $\nu_\mu$ and $\nu_\tau$ then parameter space still exists.

We note that in point number 6 where we introduce scattering experiment to help break this oscillation degeneracy, scattering experiments often come with their own degeneracies, see e.g.~\cite{Dutta:2020che,Denton:2018xmq}.
As they are quite distinct from the NSI parameters required for LMA-Dark, they are already comfortably probed by oscillation experiments.

In point number 10, we emphasize that to address the diagonal degeneracy along with the primary LMA-Dark degeneracy, we need information on $\eps_{\mu\mu}$ rather than $\eps_{\tau\tau}$.
We clarify that while $\sim\frac13$ of SNO's neutral current data set of solar neutrinos is composed of $\nu_\tau$, the deuterium scattering process is dominated by the axial-vector current which we have taken to be zero in our model to focus on the vector part which dominates for oscillations.
SNO's elastic scattering data also contains some $\nu_\tau$, but this channel is a charged current process.

One must examine very low mediator masses below which the early universe constraints may no longer apply.
Depending on the assumptions about the relative strength of the coupling between the new mediator and quarks versus neutrinos, this constraint likely weakens around $M_{Z'}\sim10$ keV based on BBN measurements \cite{Huang:2017egl}.
While this analysis is agnostic about the UV complete model to generate the NSI, it is typically expected that the coupling to neutrinos will be larger than that to quarks.
Thus, for this discussion, we conservatively assume that the quark and neutrino couplings are the same.
However, if the coupling of the mediator to the SM is sufficiently large, the new mediator will thermalize with the SM and contribute to the effective degrees of freedom of the Universe 
\cite{Escudero:2019gzq} which excludes mediators below $\sim$10 MeV. 
According to \cite{Kamada:2015era}, the constraint from $\Delta N_{\rm eff}$ measurements with the CMB rules out all mediators $M_{Z'}\lesssim5.3$ MeV. Indeed, depending on the combination of cosmological data the constraints vary between $M_{Z'}\gtrsim 3.1-10.1$ MeV at $95.4\%$ CL ($2\sigma$ at 1 d.o.f.), \cite{Sabti:2021reh} where the weakest constraint of $M_{Z'}\gtrsim 3.1$ MeV comes from BBN only.
We will conservatively use the weakest constraint only, although our narrative would not change if we used the strongest constraint cited at 10 MeV.

There is also an additional caveat to the above narrative, indicated in fig.~\ref{fig:degeneracies}.
In general the new mediator does not need to have equal couplings to up and down quarks $\eps_{\alpha\alpha}^{u,V}=\eps_{\alpha\alpha}^{d,V}$. Depending on the UV complete NSI model other combinations are possible.
To make ensure that LMA-Dark is fully ruled out all combinations of $\eps_{\alpha\alpha}^{u,V}$ and $\eps_{\alpha\alpha}^{d,V}$ need to be constrained.
Without precise measurements of oscillations in both the Sun and the Earth, we can see in fig.~~\ref{fig:dresden and osc} that a sizable allowed region will extend to the part of parameter space where $\eps_{ee}^{u,V}\sim-5$ and $\eps_{ee}^{d,V}\sim5$, even when including scattering data.
To address this, scattering data with different nuclei is needed, as the neutron fraction weakly affects the slope of the constraining region.
See also fig.~\ref{fig:future} below.

Finally, this whole narrative relies on the assumption that all experiments are consistent within a standard three flavor picture and are compatible with all $\eps_{\alpha\beta}^{f,V}=0$.
If diagonal parameters are found to be non-zero in a way that deviates from Eq.~(\ref{eq:x}) then the LMA-Dark solution is immediately lifted, although this cannot be done with oscillation data since it is a degeneracy; it could only be lifted in this way from scattering data.
Moreover, measuring one diagonal NSI parameter is not sufficient, two or three must be measured.
For the off-diagonal NSI parameters if they are non-zero and real (see e.g.~\cite{Esteban:2018ppq,Capozzi:2019iqn,Denton:2020uda,Chatterjee:2020kkm} for hints of this in long-baseline data\footnote{Interestingly, the hints for non-zero NSI in long-baseline data prefer purely complex off-diagonal values \cite{Denton:2020uda,Chatterjee:2020kkm} which do not lift the degeneracy much.}) then the entire degeneracy is also solved and additional new physics cannot imitate other scenarios with the wrong mass ordering.
This explains the slight preference from LMA over LMA-Dark in a recent global fit to oscillation data \cite{Esteban:2018ppq} as the data slightly prefers $\eps_{e\mu}\neq0$, although in \cite{Esteban:2018ppq} only  real NSI are considered.

\section{Scattering Data}
\label{sec:dresden}
Constraints on LMA-Dark come from measurements of neutrino scattering, in particular the important constraints for light mediators come from CEvNS experiments.
This process was first observed by COHERENT using neutrinos from pion decays at rest \cite{Akimov:2017ade} which disfavored LMA-Dark for mediator masses above $\sim 50$ MeV \cite{Denton:2018xmq}.
To constrain lighter mediators lower threshold detectors need to be used.
As the volume of such detectors are considerably smaller than the already fairly small COHERENT detectors, neutrinos from nuclear reactors are used to provide large statistics.
In the following we will investigate the recent measurements from the Dresden-II reactor experiment \cite{Colaresi:2021kus,Colaresi:2022obx} which presented suggestive evidence for the observation of CEvNS.

The cross section for CEvNS is given by \cite{Freedman:1973yd}
\begin{equation}
\frac{\text{d}\sigma_\alpha}{\text{d}E_R}=\frac{G_F^2}{2\pi}\frac{Q_{w\alpha}^2(q)}{4}F^2(2 M E_R)M \left( 2 -\frac{M E_R}{E_\nu^2}\right)\,,
\end{equation}
where $G_F$ is the Fermi constant, $E_R$ is the nuclear recoil energy, $F(q^2)$ is the nuclear form factor, $M$ is the mass of the target nucleus (Ge in this case), and $E_\nu$ is the incident neutrino energy. The weak charge in the presence of NSI is given by\footnote{In principle flavor changing NSI
parameters $\eps_{\alpha\beta}^{f,V}$ could also be present however we assume them to be zero as otherwise the LMA-Dark solution is not present.}
\begin{equation}
\frac{Q_{w\alpha}^2(q)}{4}=\left[Zg_p^V +Ng^V_n+3 (N+Z)\eps_{\alpha\alpha}^{f,V}(q)\right]^2\,,
\label{eq:Qw}
\end{equation}
where we have assumed that $\eps_{\alpha\alpha}^{u,V}=\eps_{\alpha\alpha}^{d,V}$ which we call $\eps_{\alpha\alpha}^{f,V}$.
Here, $N$ and $Z$ are the number of neutrons and protons in the target nucleus which we take as $(N,Z)=(40.65, 32)$ for Ge, corresponding to the weighted average
of natural isotopic abundance.
The SM couplings of the $Z$ boson to protons and neutrons at low energies are given by 
\cite{Barranco:2005yy}
\begin{align}
g_p^V&=\rho_{\nu N}^{NC}\left(\frac{1}{2}-2\kappa \hat{s}_Z^2\right)+2\lambda_{uL}+2\lambda_{uR}+\lambda_{dL}+\lambda_{dR}~,\nonumber\\
g_n^V&=-\frac{1}{2}\rho_{\nu N}^{NC}+\lambda_{uL}+\lambda_{uR}+2\lambda_{dL}+2\lambda_{dR}
\end{align}
with $\rho_{\nu N}^{NC}= 1.0082,~ \hat{s}_Z^2= \sin^2\theta_W= 0.23129,~\kappa= 0.9972,~ \lambda_{uL}=-0.0031,~\lambda_{dL}=-0.0025,$\\$\lambda_{dR}= 2\lambda_{uR}= 7.5\cdot 10^{-5}$.
Since $g_n^V\simeq-0.5$, certain values of positive $\eps_{\alpha\alpha}^{f,V}=-2(g_n^V N+g_p^V Z)/(3 (N+Z))$ lead to degenerate points with the SM \cite{Coloma:2017egw}, although these degeneracies are different from those in oscillations and are thus already constrained, as mentioned above.

The nuclear form factor $F(q^2)$ depends on the nuclear density distribution and is related to the physical size of the nucleus.
It accounts for loss of coherency at higher values of momentum transfer.
For small momentum transfers, as is the case for reactor neutrinos, $F\sim 1$ to a very good approximation.
The new physics contribution to the weak charge in Eq.~(\ref{eq:Qw}) is governed by $\eps_{\alpha\alpha}^{f,V}$\footnote{We are focusing here on vector NSI for comparison to oscillation degeneracies.
Notice that with reactor neutrinos we are only sensitive to $\alpha=e$.}
\begin{equation}
\eps_{\alpha\alpha}^{f,V}=\frac{g_{\nu_\alpha} g_f}{2\sqrt{2}G_F(q^2+M_{Z'}^2)}=\eps_{\alpha\alpha}^{f,V}(0)\frac{M_{Z'}^2}{q^2+M_{Z'}^2}\,,
\end{equation}
with $\eps_{\alpha\alpha}^{f,V}(0)$ the NSI parameter in oscillations for $q=0$ and $g_{f/\nu_\alpha}$ the coupling of the mediator to fermions and neutrinos. 

The constraints from the recent Dresden-II data on new physics scenarios have been analyzed in \cite{Liao:2022hno,Coloma:2022avw,Sierra:2022ryd,Khan:2022jnd}.
We translate the results from \cite{Sierra:2022ryd} to arrive at the excluded regions in the $\sqrt{|g_q g_{\nu_e}|}-M_{Z'}$ parameter space.
To quantify the dependence of the results on the quenching factor, the relation between nuclear recoil energy to electronic recoils, the results are derived for two different quenching factors: the first is based on a quenching factor determined using iron-filtered monochromatic neutrons provided in \cite{Colaresi:2022obx,Collar:2021fcl}, and the second is based on a modified Lindhard factor \cite{Sorensen:2014sla}.
In the following, we will always use the more conservative results which come from the modified Lindhard factor.

\begin{figure}
\centering
\includegraphics[width=\columnwidth]{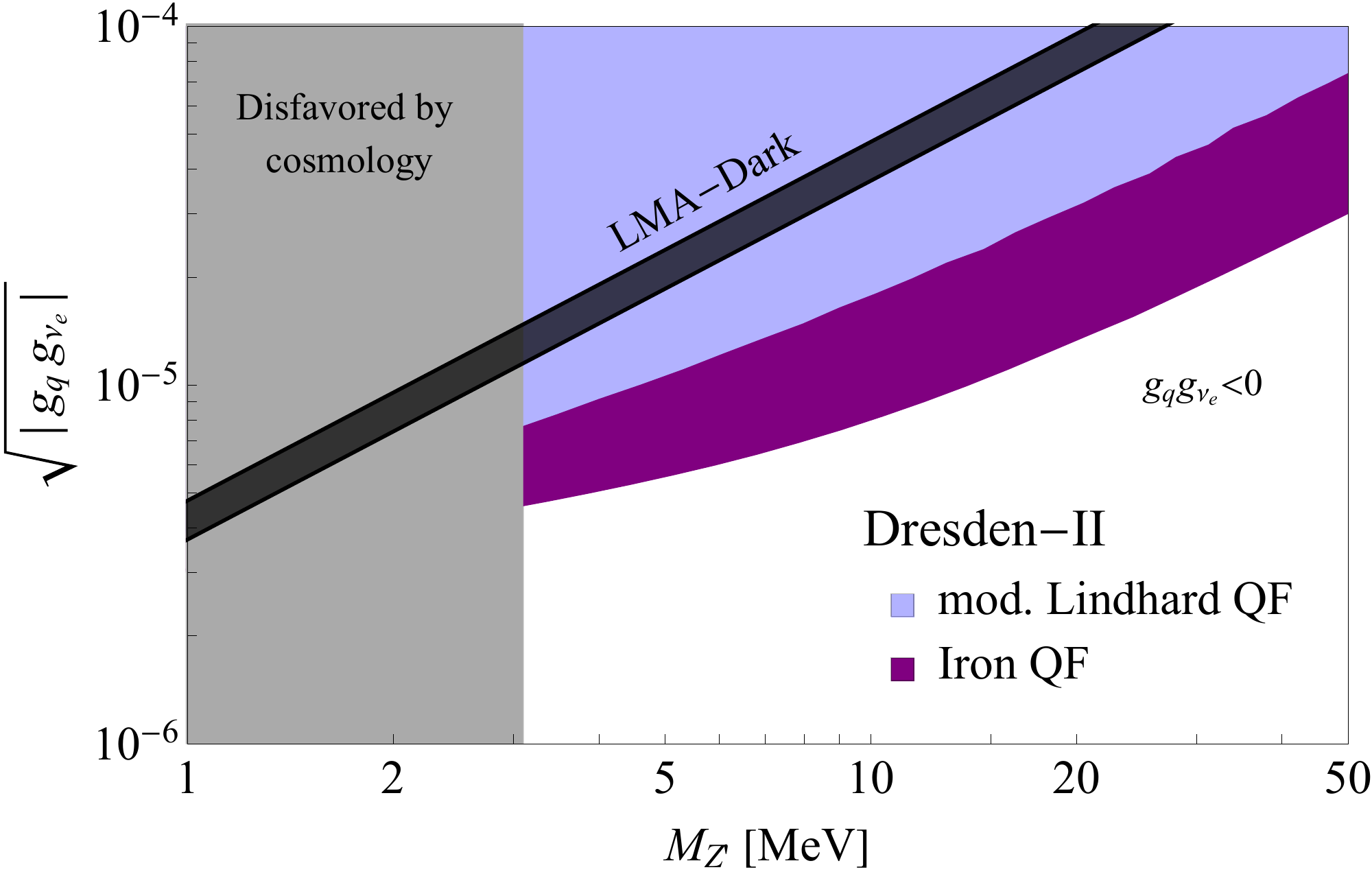}
\caption{The exclusion bounds on the coupling of the NSI mediator to $\nu_e$ and quarks over the mediator mass from Dresden-II data translated from \cite{Sierra:2022ryd}.
We require $g_{\nu_e} g_q<0$, the curves are drawn at $3\sigma$ (2 d.o.f.), and the two different shades of purple correspond to two different quenching factors.
The black band corresponds to the LMA-Dark solution to oscillation data at $3\sigma$ (1 d.o.f.) from \cite{Esteban:2018ppq}.
We assume $\eps_{ee}^{u,V}=\eps_{ee}^{d,V}$.
Additional constraints from COHERENT and other higher energy scattering experiments are only relevant for larger couplings.
The cosmology constraint is shown in gray on the left at $2\sigma$ (1 d.o.f.) \cite{Kamada:2015era}.}
\label{fig:dresden_lmad}
\end{figure}

\section{Results}
\label{sec:results}
In fig.~\ref{fig:dresden_lmad} we show the constraints from the Dresden-II data reinterpreted from \cite{Sierra:2022ryd} along with the LMA-Dark solution from \cite{Esteban:2018ppq} in the $\sqrt{|g_q g_{\nu_e}|}-M_{Z'}$ plane for $g_q g_{\nu_e}<0$.
We see that independent of the quenching factor, the LMA-Dark solution is excluded by the Dresden-II data at more than $3\sigma$ (2 d.o.f.).

\begin{figure}
\centering
\includegraphics[width=\columnwidth]{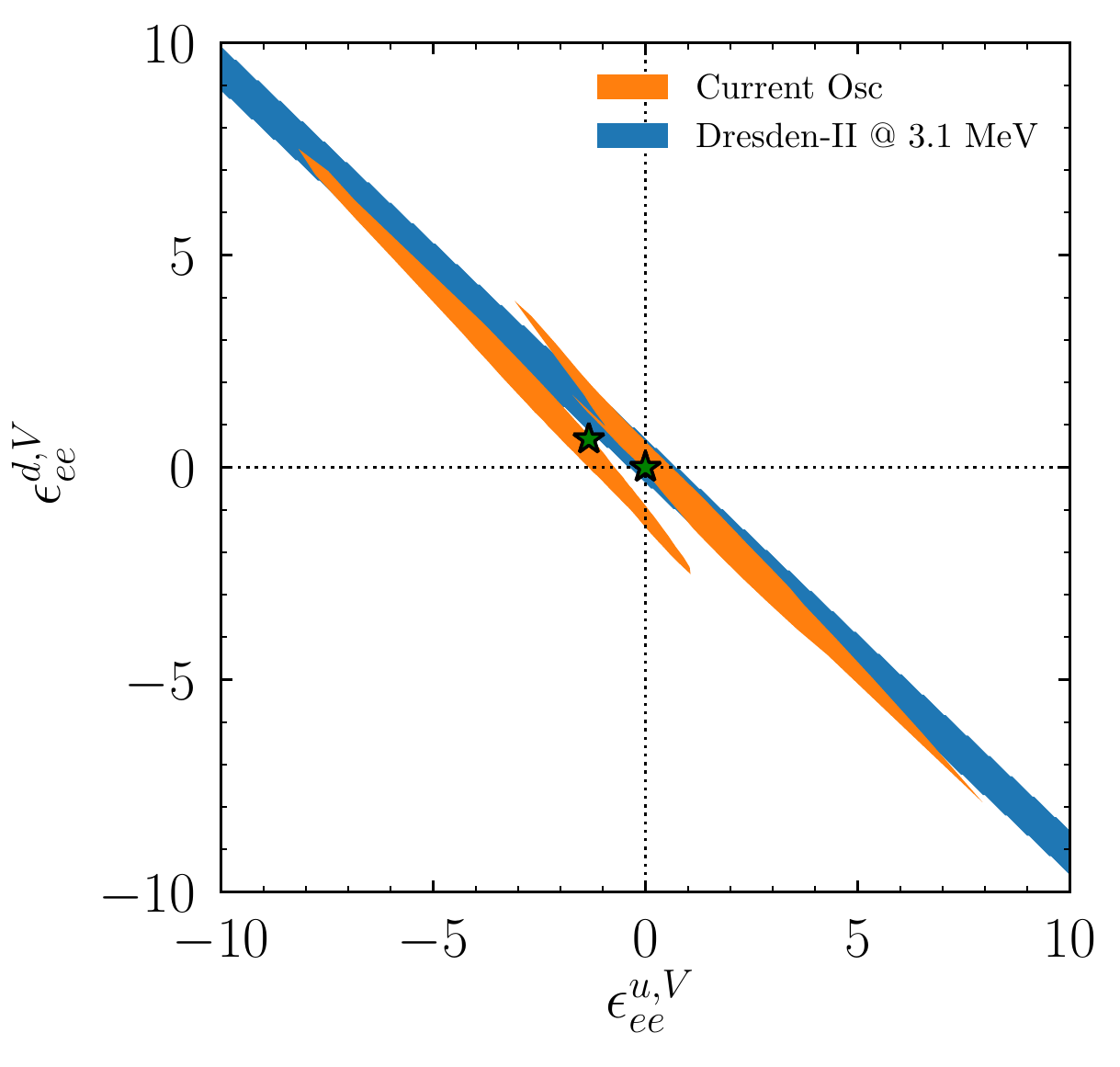}
\caption{The same as fig.~\ref{fig:degeneracies} 
including the allowed regions from oscillations at $3\sigma$ (1 d.o.f.) adapted from \cite{Esteban:2018ppq} in orange, see text for details. 
In blue we show the constraints calculated in this paper from the Dresden-II data at $3\sigma$ (1 d.o.f.) for mediators with mass $M_{Z'}=3.1$ MeV, the lower limit from cosmology.}
\label{fig:dresden and osc}
\end{figure}

In fig.~\ref{fig:dresden and osc} we show the preferred regions from the Dresden-II data in the $\eps_{ee}^{u,V}-\eps_{ee}^{d,V}$ plane along with the preferred region from a global fit to oscillations, adapted from \cite{Esteban:2018ppq}.
For the Dresden-II region, we fix the mediator mass to $M_{Z'}=3.1$ MeV corresponding to the lightest mediator mass allowed by cosmology \cite{Sabti:2021reh}. 
We also briefly comment on the oscillation global fit preferred regions.
First, we note that for sufficiently large values of $|\eps_{ee}^{q,V}|$, both the LMA and LMA-Dark regions from oscillations taper off and are closed.
Second, we see various spikes corresponding to different slopes coming from different neutron fractions.
The spike at the Earth's neutron fraction is due to a small amount of information of the matter effect coming from NOvA and T2K at two different densities, see \cite{Bharti:2020gnu}, but mostly from the day-night effect of solar neutrinos in the Earth combined with KamLAND measurements in (near) vacuum.
There is also some information about the matter effect coming from combining regular solar neutrino measurements and KamLAND at two different neutron fractions, see fig.~\ref{fig:degeneracies}.
Finally, we comment on the spikes in the LMA region in the $\eps_{ee}^{d,V}>0$ -- $\eps_{ee}^{u,V}<0$ quadrant.
By comparing fig.~\ref{fig:dresden and osc} with fig.~\ref{fig:degeneracies}, we can see that these spikes occur at the the slopes for the Earth and for the Sun.
Thus, in that direction as well, we can again see clearly that there is some information on the matter effect coming from both experiments in the Sun and in the Earth.
We also note that both regions are not centered exactly on the SM and the fully degenerate point; this is because of the slight tension between solar neutrino data and KamLAND in $\Delta m^2_{21}$\footnote{Newer data from Super-KamiokaNDE somewhat alleviates this tension, but is not included in the fit \cite{Esteban:2020cvm} that went into fig.~\ref{fig:dresden and osc}.}.
This solar tension can be explained with $\eps_{ee}\sim0.1$ in the Sun \cite{Liao:2017awz} which is both far away and in the opposite direction from the NSI required for LMA-Dark of $\eps_{ee}=-2$.

\begin{figure}
\centering
\includegraphics[width=\columnwidth]{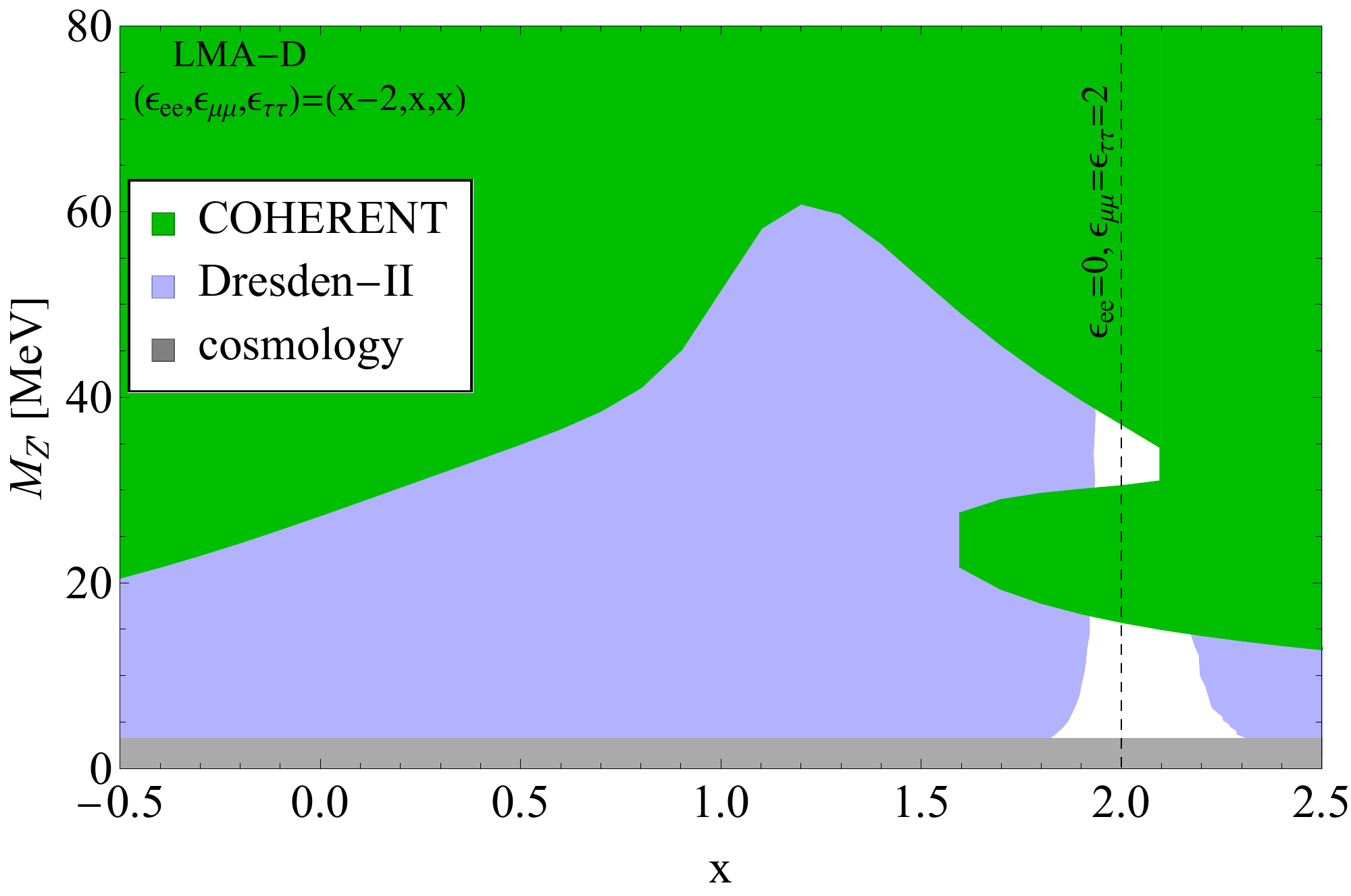}
\caption{Constraints on $x$ which parameterizes whether LMA-Dark is in $\nu_e$ sector ($x=0$) or the $\nu_\mu$ and $\nu_\tau$ sector ($x=2$), see Eq.~(\ref{eq:x}).
The COHERENT (green, CsI data reanalyzed from \cite{Denton:2018xmq}) and Dresden-II (blue, derived in this paper) constraints are at $3\sigma$ (2 d.o.f.). The gray region shows the mediator masses ruled out by cosmology at $2\sigma$ (1 d.o.f.) \cite{Kamada:2015era}.}
\label{fig:Xmz}
\end{figure}

\begin{figure}
\centering
\includegraphics[width=\columnwidth]{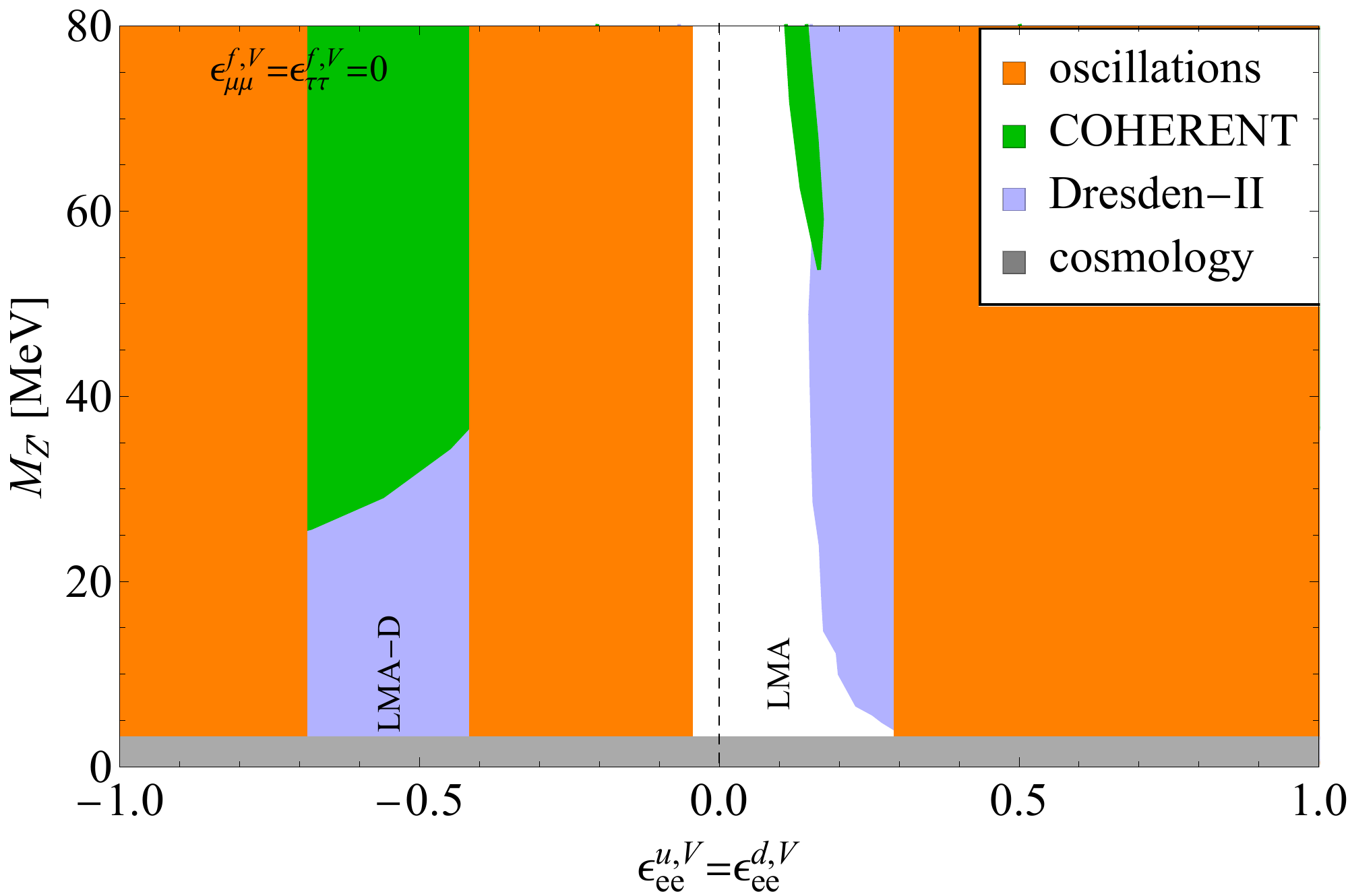}
\caption{Constraints on $\epsilon_{ee}^{u,V}=\epsilon_{ee}^{d,V}$ and the mediator mass $M_{Z'}$ at $3\sigma$ (2 d.o.f.), assuming $\epsilon_{\mu\mu}^{f,V}=\epsilon_{\tau\tau}^{f,V}=0$. The blue regions are excluded by the Dresden-II data, the green regions by COHERENT-CsI (reanalyzed from \cite{Denton:2018xmq}), the gray region by cosmology. The orange regions are excluded from oscillation data which leave only the LMA and LMA-Dark regions as allowed parameter space. The COHERENT constraints for $\epsilon_{ee}^{f,V}>0$ are weaker than the oscillation constraints and are therefore covered by them in the plot.}
\label{fig:epseemz}
\end{figure}

\begin{figure}
\centering
\includegraphics[width=\columnwidth]{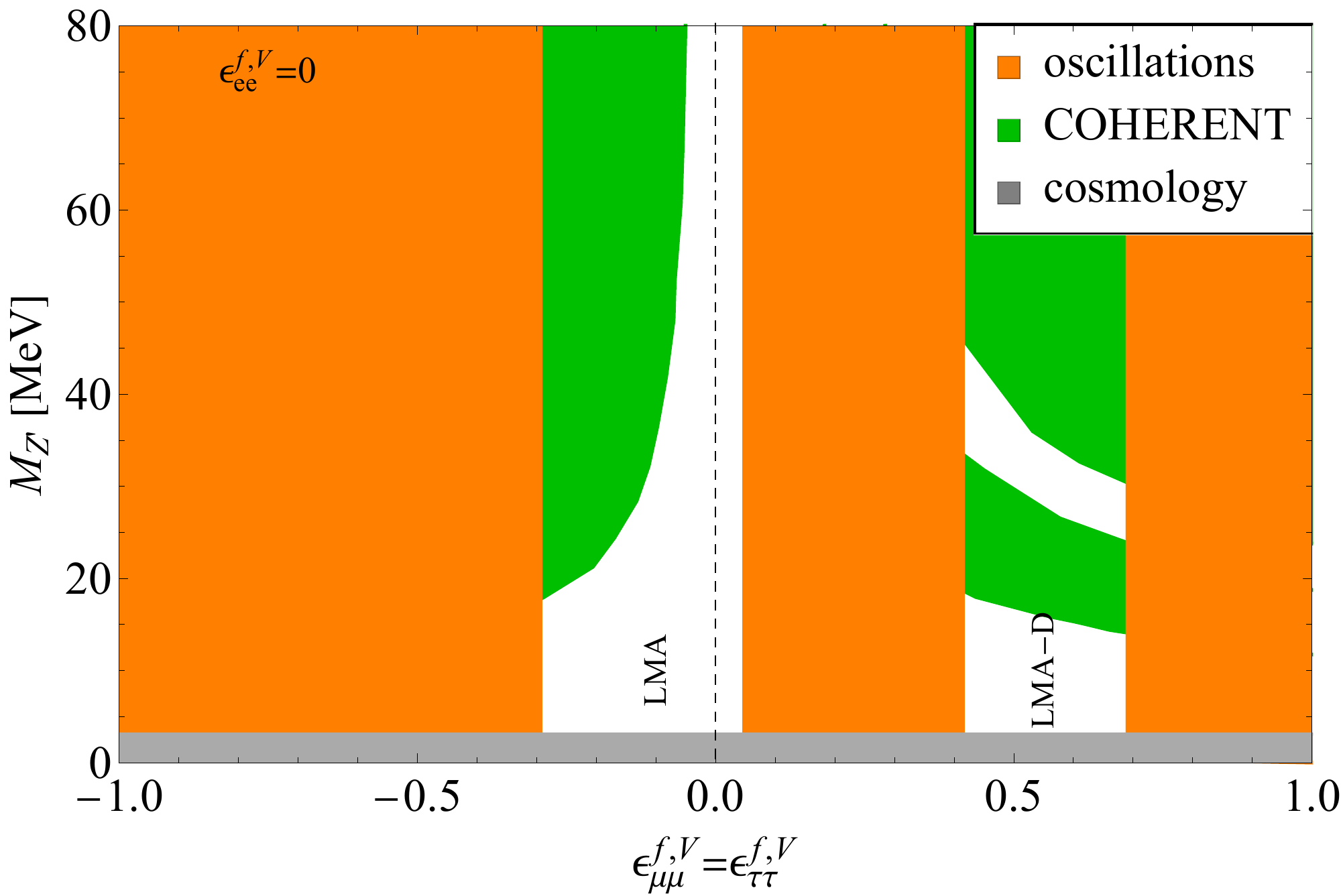}
\caption{The same as fig.~\ref{fig:epseemz}, but for $\epsilon_{\mu\mu}^{f,V}=\epsilon_{\tau\tau}^{f,V}$, assuming $\epsilon_{ee}^{f,V}=0$.
As there is no constraint on $\epsilon_{\mu\mu}^{f,V}$ or $\eps_{\tau\tau}^{f,V}$ for mediator masses below 50 MeV, parts of the LMA-Dark region are still allowed, indicated by the white regions on the right of the figure.}
\label{fig:epsmmmz}
\end{figure}

From fig.~\ref{fig:dresden_lmad} we see that with the Dresden-II data the LMA-Dark solution in the $\nu_e$ sector only is ruled out for equal up and down quark couplings.
This, however, leaves the possibility open that LMA-Dark is realized in the $\nu_{\mu}$ and $\nu_\tau$ sectors. To quantify how much of the LMA-Dark degeneracy is in the $\nu_e$ sector versus the $\nu_\mu$ and $\nu_\tau$ sector, we use the fact that the SM oscillation probability in matter is unchanged by any contribution to the Hamiltonian proportional to the identity matrix, \cite{Denton:2018xmq}, see Eq.~(\ref{eq:x}) and zero off-diagonals.
In fig.~\ref{fig:Xmz} we show the constraint on $x$ as a function of the mediator mass coming from COHERENT-CsI data \cite{Denton:2018xmq}, the Dresden-II data, and the constraint on the mediator mass from cosmology.
This shows the picture if oscillations measured exactly the standard picture with arbitrary precision in both the Earth and the Sun.
While COHERENT rules out mediator masses $M_{Z'}\gtrsim 20-60$ MeV, the Dresden-II constraints extend to lower mediator masses such that $x\lesssim 1.9$ and $x\gtrsim 2.2$ are excluded for all mediator masses. This leaves only the narrow region of $x\in[1.9,2.2]$, where the LMA-Dark solution is mainly in the muon and tau sector instead of in the electron sector,
for light mediator masses $M_{Z'}\in[3.1,15]$ MeV which are not probed yet and the mediator masses for which COHERENT is not constraining due to degeneracies in the CEvNS cross section around $M_{Z'}\in[30,40]$ MeV.
In fact, from fig.~\ref{fig:epseemz} we see that for $\eps_{ee}^{f,V}$ with $\eps_{\mu\mu}^{f,V}=\eps_{\tau\tau}^{f,V}=0$ the LMA-Dark region is fully excluded with Dresden-II data, which only leavess the allowed region between $\eps_{ee}^{f,V}\in[-0.05,0.2]$
where the constraints from scattering data is even more constraining than the oscillation constraints alone. The situation is different assuming $\eps_{ee}^{f,V}=0$ and $\eps_{\mu\mu}^{f,V}=\eps_{\tau\tau}^{f,V}\neq 0$
shown in fig.~\ref{fig:epsmmmz}.
In this case, no constraints from Dresden-II apply, so that generally only mediator masses above 50 MeV are constrained by the COHERENT-CsI data.
This leads to parts of the LMA-Dark solution still being allowed for $M_{Z'}\lesssim 50$ MeV in the region where the effects of NSI in the scattering cross section are degenerate with the SM and below $M_{Z'}\lesssim 20 $ MeV where COHERENT loses sensitivity. 
In fact, even with hypothetical high precision COHERENT data, the lower green band on the right of fig.~\ref{fig:epsmmmz} would stretch down, but the gap between the two bands would persist since the couplings $g_\nu$, $g_q$ have the same sign; this region could be covered with different target nuclei with significantly different neutron fractions.
Nevertheless, for the LMA region and $M_{Z'}\gtrsim 20 $ MeV scattering data leads to tighter constraints on the NSI parameters than oscillation data only.

To summarize our results, with the latest Dresden-II data, there are now only two ways to achieve the LMA-Dark degeneracy which allows for an incorrect mass ordering determination:
\begin{itemize}
\item If the new physics is in the $\nu_e$ sector via $\eps_{ee}$, with very specific couplings to up and down quarks given by the overlapping regions in fig.~\ref{fig:dresden and osc}.
In this case, the mediator must be lighter than $\sim50$ MeV as COHERENT has already ruled out heavier mediator masses for all combinations of up and down quark couplings \cite{Denton:2018xmq,Chaves:2021pey}.
\item If the new physics is in the $\nu_\mu$ and $\nu_\tau$ sectors with similar or equal couplings.
In this case, the mediator needs to be lighter than $\sim40$ MeV for $\eps_{\mu\mu}^{u,V}=\eps_{\mu\mu}^{d,V}=\eps_{\tau\tau}^{u,V}=\eps_{\tau\tau}^{d,V}$, see fig.~\ref{fig:epsmmmz}.
For some combinations of up and down quark combinations, NuTeV data for $M_{Z'}\gtrsim10$ GeV provides the only constraint \cite{Coloma:2017egw}.
\end{itemize} 

\section{The Future}
\label{sec:future}
In the future, improvement on resolving the LMA-Dark degeneracy will come on the oscillation side by improving the constraint on $\eps_{ee}$.
This will predominantly come from improvements in the comparison of $\Delta m^2_{21}$ between vacuum measurements (by KamLAND \cite{Gando:2013nba} and JUNO \cite{Djurcic:2015vqa} in the future) and solar experiments (SNO \cite{Ahmad:2002jz}, Borexino \cite{BOREXINO:2018ohr}, and SK \cite{Super-Kamiokande:2016yck}, as well as DUNE in the future \cite{Capozzi:2018dat,DUNE:2020ypp}).
As the measurement by JUNO will be extremely precise, we focus on the precision of the solar measurement from DUNE.
To a very good approximation, the precision on $\eps_{ee}$ in the Sun is the same as the level of agreement on $\Delta m^2_{21}$.
From \cite{Capozzi:2018dat}, DUNE will measure $\Delta m^2_{21}$ with a precision of 6\% while JUNO will measure $\Delta m^2_{21}$ much better at the sub-percent level \cite{Djurcic:2015vqa}.
As DUNE's measurement comes predominantly from the day-night asymmetry, the relevant neutron fraction is that for the Earth, $Y_n\sim1.05$.

\begin{figure}
\centering
\includegraphics[width=\columnwidth]{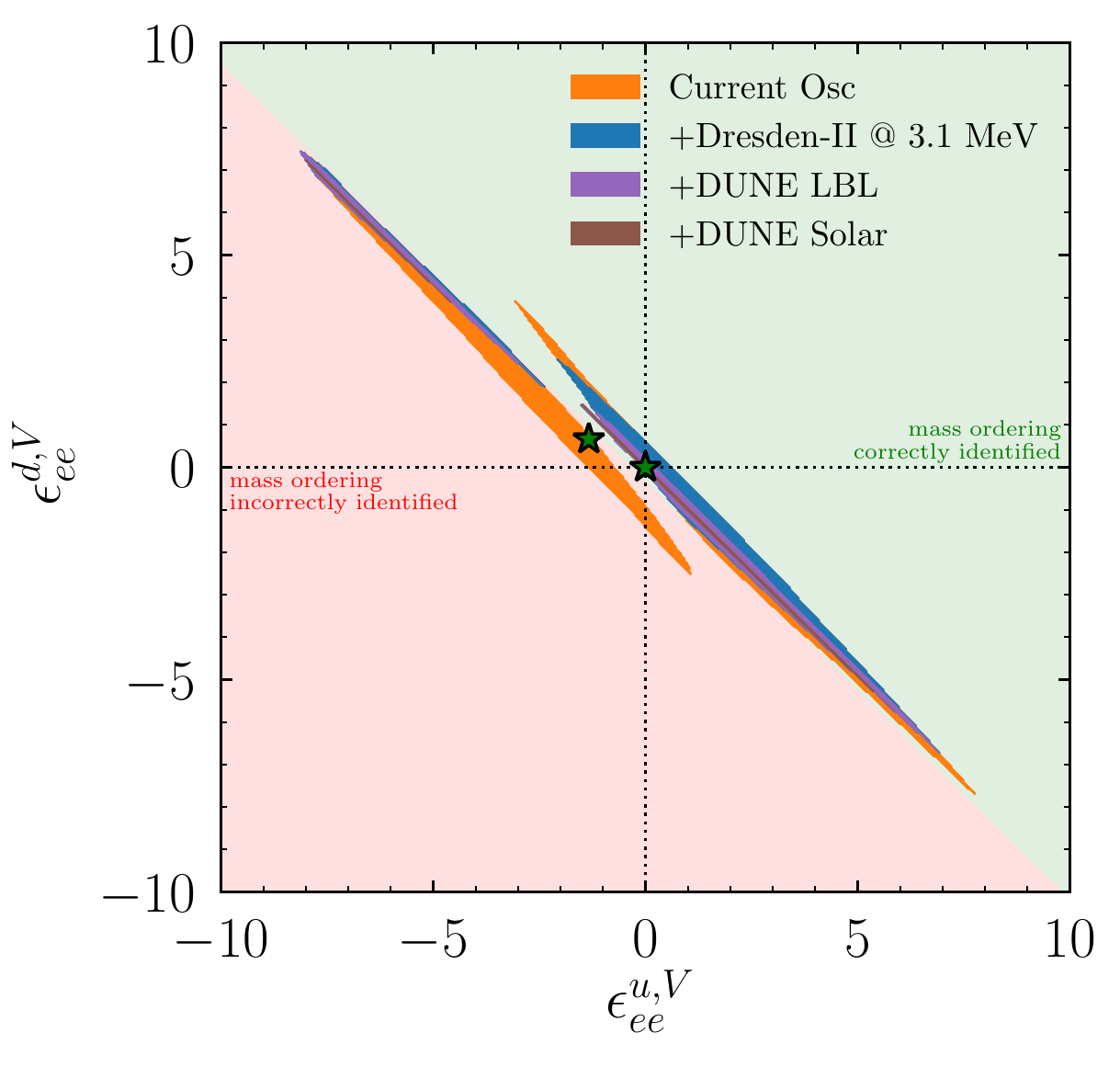}
\caption{The same as figs.~\ref{fig:degeneracies}, \ref{fig:dresden and osc} at $M_{Z'}=3.1$ MeV but with anticipated improvements from future measurements of oscillation data from DUNE, see text for details. The red region in the lower left half-plane shows the values of NSI parameters where the mass ordering is incorrectly identified and the green region shows where the mass ordering can be correctly identified.
Note that even with future measurements of the matter effect from DUNE solar and long-baseline accelerator data, a small brown sliver still exists in the LMA-Dark region which is ruled out for much heavier mediators \cite{Coloma:2016gei} by CHARM data \cite{CHARM:1986vuz}, not shown.}
\label{fig:future}
\end{figure}

In addition, DUNE can measure the density of the Earth with $\sim25\%$ precision at $1\sigma$ \cite{Kelly:2018kmb} which means that DUNE will measure $\eps_{ee}$ with $\sim25\%$ precision.
The effect of this and DUNE's solar measurement are shown in fig.~\ref{fig:future}.
DUNE also has some sensitivity to $\eps_{ee}$ in the Earth \cite{Denton:2021rgt} via low energy atmospheric neutrinos \cite{Kelly:2019itm,DUNE:2020ypp}.

Finally, combining measurements of CEvNS in materials with different neutron fractions \cite{Akimov:2018ghi,Denton:2018xmq,Giunti:2019xpr,Akimov:2019rhz,Barranco:2005yy,Dent:2017mpr} will also help in lifting the degeneracy as the slope of the CEvNS constraints in the $\eps_{ee}^{u,V}-\eps_{ee}^{d,V}$ plane depends on the atomic number and the number of neutrons of the detector material as $(2N+Z)/(2Z+N)$. This will lead to a narrowing of the allowed regions from scattering experiments.
Depending on the precision of these experiments and the exact targets used, this can rule out the remaining sliver in the top left region of parameter space of fig.~\ref{fig:future} for certain combinations of up and down quarks.
For heavy mediators $\gtrsim10$ GeV, data from CHARM \cite{CHARM:1986vuz}, in combination with oscillation data and Dresden-II, rules out the LMA-Dark values for any combination of up and down quark couplings \cite{Coloma:2016gei,Coloma:2017egw}.
We note that the best experiments to probe the remaining region of parameter space for lighter mediators are CEvNS experiments with small neutron fractions and thus smaller atoms, although these do not benefit from the $N^2$ enhancement of CEvNS as much making such a detection more challenging.
Possible experiments for this are the NUCLEUS experiment using silicon and germanium detectors at a reactor \cite{NUCLEUS:2019igx} or a silicon detector at the ESS \cite{Baxter:2019mcx}.

To be able to fully rule out the LMA-Dark solution and 
guarantee that upcoming oscillation experiments can unambiguously measure the mass ordering LMA-Dark also needs to be ruled out in the $\nu_\mu$ and $\nu_\tau$ sector as well as in the $\nu_e$ sector for very specific values of the up and down quark couplings.
Future low-threshold $\pi$-DAR experiments like Coherent CAPTAIN Mills \cite{CCM:2021leg} or at the ESS \cite{Baxter:2019mcx,Chaves:2021pey} can probe $\eps_{\mu\mu}$ down to low mediator masses and the ESS could feature a silicon detector with a favorable (small) neutron fraction to address the specific couplings to up and down quarks.
Also, multi-ton scale Dark Matter experiments can probe NSI as it affects the neutrino floor \cite{AristizabalSierra:2017joc}.
Even though we cannot probe $\eps_{\tau\tau}$ directly with low threshold scattering experiments \cite{Abraham:2022jse}\footnote{An exception is CEvNS with solar neutrinos, as the neutrinos leave the Sun in a mass eigenstate which contains $\nu_\tau$.}, the requirement of the LMA-Dark solution for equal $\eps_{\mu\mu}$ and $\eps_{\tau\tau}$ means that constraining only $\eps_{\mu\mu}$ is sufficient.

\section{Conclusions}
\label{sec:conclusions}
Measuring the neutrino mass ordering is one of the main goals of upcoming neutrino experiments. To ensure that this measurement can be unambiguously interpreted, any new physics effects need to be excluded. In particular, oscillation experiments suffer from the LMA-Dark degeneracy where the simultaneous change of the sign of the atmospheric mass splitting and the introduction of new neutrino interactions leave the neutrino evolution invariant.

Scattering experiments, in particular CEvNSs experiments, break this degeneracy and provide valuable insights into the existence of new neutrino interactions. However, in general, they are only sensitive to mediator masses which exceed the momentum transfer of the process which up to now restricted the constraints to mediator masses to above 50 MeV. In this paper we have made use of the first CEvNS data using reactor neutrinos which allows us to place constraints on lower mediator masses down to the few MeV level, below which constraints from the early Universe apply. With these results we have ruled out the LMA-Dark solution in the electron neutrino sector except for very specific combinations of up and down quark couplings.
This means that the only possibility for which oscillation experiments cannot unambiguously measure the mass ordering is due to new physics in the $\nu_\mu$ and $\nu_\tau$ sector with a mediator mass in the small region between $\sim3$ and $\sim50$ MeV coupling to electrons or specific up and down quark couplings, or for new physics in the $\nu_e$ sector with very specific couplings to up and down quarks.
These region can be probed by the upcoming Coherent CAPTAIN Mills experiment or a CEvNS detector at the European Spallation Source which benefit from having $\nu_\mu$'s and also a different neutron fraction from Dresden-II.
Our results therefore solidify the possibility for a robust determination of the neutrino mass ordering by upcoming oscillation experiments such as DUNE, JUNO, and others.

\begin{acknowledgments}
We thank Valentina De Romeri for sharing their Dresden-II results and for useful discussions about the simulation of the Dresden-II data.
We acknowledge support from the US Department of Energy under Grant Contract DE-SC0012704.
Some figures were prepared with \texttt{python} \cite{10.5555/1593511} and \texttt{matplotlib} \cite{Hunter:2007}.
\end{acknowledgments}

\appendix
\section*{Appendix: The Necessity of Solar Data for JUNO's Atmospheric Mass Ordering Determination}
\label{sec:JUNO}
JUNO will have sensitivity to the atmospheric mass ordering.
However regardless of the precision of the experiment which depends on subtle experimental details, input from solar experiments is required for a determination of the atmospheric mass ordering.
This argument follows from two distinct steps, both involving different measurements that will come from JUNO (one of which also came from KamLAND), as well as data from solar experiments.

First, JUNO needs to know if $\theta_{12}<45^\circ$ or $>45^\circ$\footnote{Alternatively, depending on one's definition of $|\nu_1\rangle$ and $|\nu_2\rangle$, this is the statement that JUNO needs to know if $\Delta m^2_{21}$ is positive or negative.}.
This can be see theoretically in that measurements of solar parameters in reactor experiments such as KamLAND and JUNO are only sensitive to $\sin^22\theta_{12}$ and thus cannot differentiate between these two cases.
Meanwhile, $^8$B solar data from SNO, SK, and Borexino, as well as other experiments such as Homestake, is sensitive to $\sin^2\theta_{12}$ and thus tells us which octant $\theta_{12}$ is in, in particular, that it is in the lower octant.

Second, knowing the octant of the solar mixing angle is necessary to determine the atmospheric mass ordering at JUNO.
This is because JUNO will observe both atmospheric frequencies $\Delta m^2_{31}$ and $\Delta m^2_{32}$ and determine which is larger, thus telling us which mass ordering is correct.
Identification of the two mass orderings is only possible because the prefactor governing the size of the oscillations is proportional to $s_{12}^2\sim1/3$ for $\Delta m^2_{32}$ and $c_{12}^2\sim2/3$ for $\Delta m^2_{31}$.
If we only knew the value of $\sin^22\theta_{12}$, which is the case without solar data, then we could swap $c_{12}^2$ and $s_{12}^2$ and still get the same fit to reactor data.
This statement is exact at the probability level thus is independent of any experimental details.

\begin{figure}
\centering
\includegraphics[width=\columnwidth]{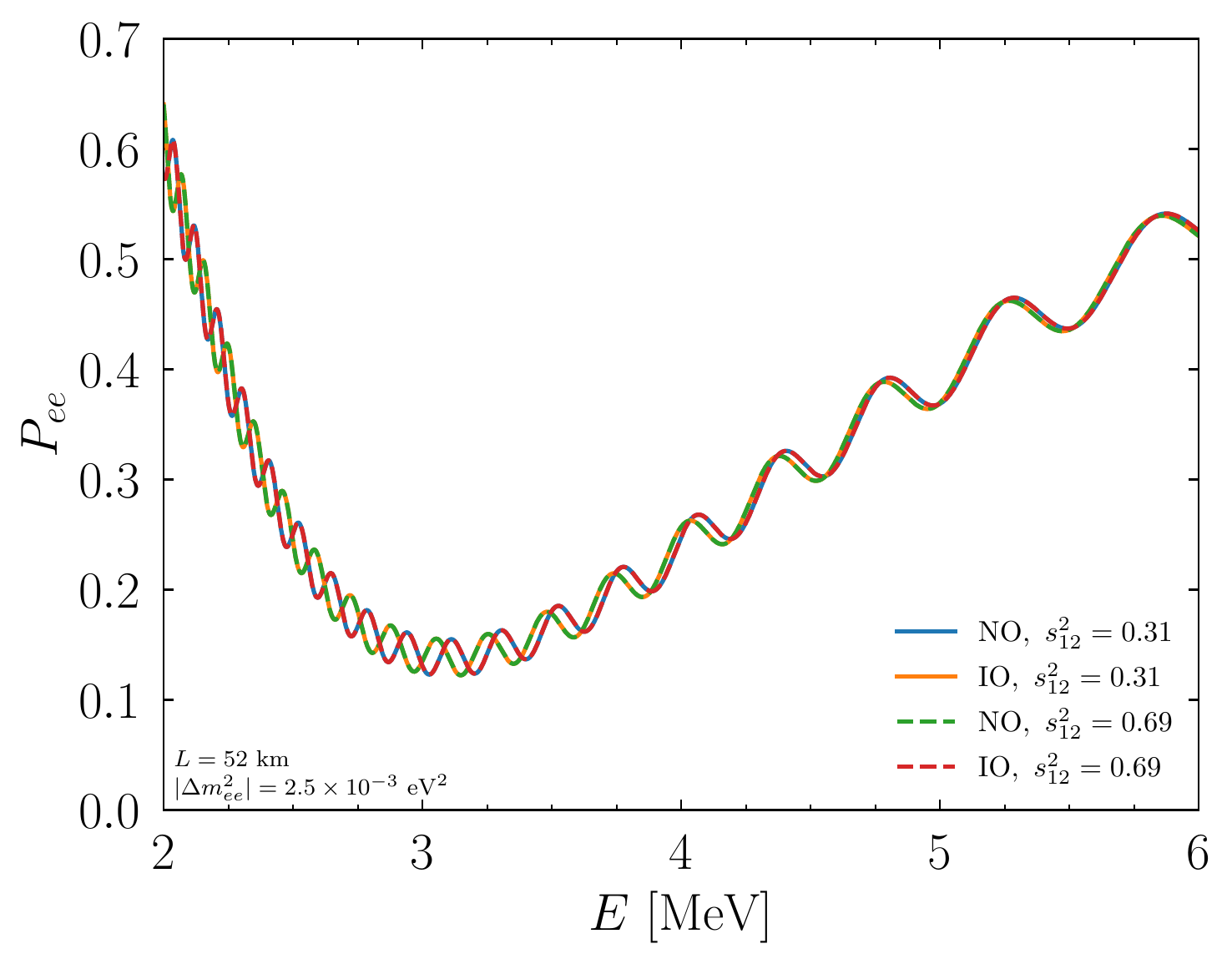}
\caption{The disappearance probability for JUNO-like parameters.
We see that if we do not know that octant of $\theta_{12}$, then both atmospheric mass orderings are truly identical regardless of the experimental sensitivity.
Since solar data tells us that $s_{12}^2\sim0.31$, we know that we are in either the blue or the orange case and thus JUNO can differentiate between the two mass orderings.}
\label{fig:Pee JUNO}
\end{figure}

This is explicitly shown numerically in a full 3-flavor calculation in fig.~\ref{fig:Pee JUNO}.

\bibliography{main}

\end{document}